\documentstyle[preprint,psfig,aps]{revtex}
\draft
             
\begin{document}                

\def\be{\begin{equation}}
\def\ee{\end{equation}}
\def\ba{\begin{eqnarray}}
\def\ea{\end{eqnarray}}
\def\ban{\begin{eqnarray*}}
\def\ean{\end{eqnarray*}}


\title{Exact solutions for interacting boson systems under rotation}
\author{Thomas Papenbrock\cite{tp} and George F. Bertsch\cite{gfb}}
\address{Institute for Nuclear Theory, Department of Physics, 
University of Washington, Seattle, WA 98195, USA}
\maketitle
\begin{abstract}
We study a class of interacting, harmonically trapped boson systems at 
angular momentum $L$. The Hamiltonian leaves a $L$-dimensional
subspace invariant, and this permits an explicit solution of
several eigenstates and energies for a wide class of two-body interactions. 
\end{abstract}
The study of vortices in Bose-Einstein condensates of trapped atomic vapors is
of considerable experimental \cite{JILA,Dalibard} and theoretical
\cite{Lundh,Rokhsar,Holland,Feder,Fetter,Linn,Wilkin,Mottelson,Bertsch,Kavoulakis,Jackson}
interest.  Present experiments and much theoretical work
\cite{Lundh,Rokhsar,Holland,Feder,Fetter} focus on the Thomas-Fermi regime of
short coherence length \cite{Stringari}.  The opposite limit of perturbatively
weak contact interactions is not yet realized experimentally but is of
theoretical interest as well
\cite{Linn,Wilkin,Mottelson,Bertsch,Kavoulakis,Jackson}. The case of
attractive interactions was studied by Wilkin {\it et al.}\cite{Wilkin}.
Mottelson developed a theory for repulsive interactions \cite{Mottelson}, and
Kavoulakis {\it et al.}  \cite{Kavoulakis} and Jackson {\it et al.}
\cite{Jackson} compared mean field and exact numerical results. Exact
diagonalization techniques by us showed that the ground state energy depends
linearly on angular momentum and led to an analytical expression for the ground
state wave function \cite{Bertsch}. Recently, the form of this wave function
and its eigenenergy were confirmed in analytic calculations
\cite{Jackson2,Smith,PapBert}. However, we still lack insight about why certain
eigenstates turn out to be simple analytical functions.

We show in this paper that there is a subspace structure that explains
these findings. Interestingly, this structure is not limited to the 
two-body contact interaction considered in previous investigations but 
present in a rather large class of Hamiltonians. 
Let us consider a
system of $N$ harmonically trapped bosons in two spatial dimensions
\footnote{The three dimensional problem is essentially two-dimensional, see
e.g. ref.\cite{Jackson}} at angular momentum $L$. The trap Hamiltonian is
\be
\label{trap}
\hat{H}={1\over 2}\sum_{j=1}^N\left(-\nabla_j^2 + r_j^2 - 2\right),
\ee
where $r_j^2=x_j^2+y_j^2$. Let us use complex single-particle coordinates
$z_j=x_j+iy_j, \quad j=1\ldots N$ and fix the angular momentum to be $L$. 
The eigenfunctions of the trap Hamiltonian are of the form
\be
\label{wave}
\psi(z_1,\ldots,z_N)=\phi(z_1,\ldots,z_N)\exp{\left(-{1\over
2}\sum_{j=1}^N |z_j|^2\right)}
\ee
and have degenerate energy $E=L$. In eq.(\ref{wave}) 
$\phi(z_1,\ldots,z_N)$ denotes a homogeneous polynomial of degree $L$ that is
totally symmetric under permutation of particle indices. Suitable basis
functions for such polynomials are products 
\ban
\phi(z_1,\ldots,z_N)=e_{\lambda_1} e_{\lambda_2} \ldots e_{\lambda_k}
\ean
where $e_\lambda$ denotes the elementary symmetric polynomial \cite{Macdonald}
\ban
e_\lambda(z_1,\ldots,z_N)=\sum_{1\le p_1<p_2<\ldots<p_\lambda\le N} z_{p_1} 
z_{p_2}\ldots z_{p_\lambda},
\ean
and $\{\lambda_1,\lambda_2,\ldots,\lambda_k\}$ is a partition of $L$ into at
most $N$ integers, e.g. $\sum_{j=1}^k\lambda_j = L$ with $\lambda_1 \ge
\lambda_2 \ge \ldots \ge \lambda_k\ge 0$ and $k\le N$. We set $e_0=1$.  From
here on we omit the notation of the ubiquitous exponentials in the wave
function and restrict ourselves to the regime $0\le L\le N$. In this regime,
$N$ is simply a parameter of the problem, and the dimension of Hilbert space
equals the number of partitions of $L$ into integers.

The operator for the total angular momentum is 
\be
\label{L} 
\hat{L}=\sum_{j=1}^N z_j\partial_j,
\ee 
where $\partial_j$ denotes the derivative with respect to $z_j$ and acts 
only onto the polynomial part of the wave function.
Another important observable is the angular momentum of the center of mass.
Its operator is given by
\be
\label{Lc}
\hat{L}_c\equiv z_c\,D_c,
\ee
where $z_c=e_1/N$ denotes the center of mass and
\be
\label{Dc}
D_c=\sum_{j=1}^N\partial_j.
\ee 
Again, it is understood that derivatives act on the polynomial part of the
wave function only. $L_c$ has eigenvalues
$0,1,2,\ldots,L-2,L$ and commutes with the harmonic trap Hamiltonian 
(\ref{trap}) and the angular momentum (\ref{L}). We further have the following
commutator relations
\ba
\label{raise}
[\hat{L},D_c]=-D_c, \quad [\hat{L},z_c]=z_c,\quad[\hat{L}_c,D_c]=-D_c, 
\quad [\hat{L}_c,z_c]=z_c.
\ea

The introduction of perturbatively weak
interactions lifts the degeneracy of the trap Hamiltonian. In what 
follows we are interested in two-body interactions of the form
\ba
\label{ham}
\hat{V}&=g&\sum_{m=0}^L c_m \hat{A}_m, \nonumber\\
\hat{A}_m&=&\sum_{1\le i<j\le N} \left(z_i-z_j\right)^m\,\left(\partial_i - 
\partial_j\right)^m,
\ea
where the derivatives act only onto the polynomial part of the wave
function. We choose $gN\ll 1$ to be in the perturbative regime. 
Note that a large class of two-body interactions is of the form (\ref{ham}).
Consider for instance the two-body potential $V(r)=r^{2n}$. Comparison of 
matrix elements shows that
\ban
\left\langle\sum_{i<j} (r_i-r_j)^{2n}\right\rangle
&=&\left\langle\sum_{i<j}\left[(z_i^*-z_j^*)(z_i-z_j)\right]^n\right\rangle\\
&=&\left\langle\sum_{i<j}(\partial_i-\partial_j)^n(z_i-z_j)^n\right\rangle\\
&=&\left\langle\sum_{m=0}^n {n\choose m}{n!\over m!} 
2^{n-m}\,\hat{A}_m\right\rangle.
\ean
Thus, any potential that depends analytically on the squared inter-particle
distance $r^2$ can be written in the form of eq.~(\ref{ham}). Further examples
include zero range potentials of the form 
\ban
V(r)=2\pi g \left[a_0\delta(r) + a_1 \nabla^2\delta(r) + 
a_2\nabla^4\delta(r)\right].
\ean
For $c_m=(-1/2)^m/m!$ the interaction (\ref{ham}) corresponds to the two-body
contact interaction $V=2\pi g\delta(\vec{r})$.  This can be seen by computing
and comparing matrix elements of these two different representations of the
interaction. It is also instructive to analyze the action of the operator
(\ref{ham}) on the wave function. Inserting the appropriate coefficients for
the contact interaction into
the Hamiltonian (\ref{ham}) we obtain a Taylor series. Thus,
\ban
\lefteqn{\sum_{m=0}^L {1\over m!} 
\left({z_j-z_i\over 2}\right)^m\,\left(\partial_i -
\partial_j\right)^m \phi(z_1,\ldots,z_i,\ldots,z_j,\ldots,z_N)}&&\nonumber\\
&=&\sum_{m,n=0}^L {(-1)^n\over m!n!}\left({z_j-z_i\over 2}\right)^{m+n}
\partial_i^m\partial_j^n\,\phi(z_1,\ldots,z_i,\ldots,z_j,\ldots,z_N)\nonumber\\
&=&\phi\left(z_1,\ldots,z_i-{z_i-z_j\over2},\ldots,
z_j-{z_j-z_i\over 2},\ldots,z_N\right)\nonumber\\
&=& \phi\left(z_1,\ldots,{z_i+z_j\over2},\ldots,
{z_j+z_i\over 2},\ldots,z_N\right).
\ean
In ref.\cite{PapBert} we presented an alternative differential operator with 
this effect on wave functions. 

The operators $\hat{A}_m$ commute with $\hat{L},\hat{L}_c, z_c$ and
$D_c$. Eq.(\ref{raise}) thus implies that the application of $z_c$ to
eigenstates with with quantum numbers $L,L_c$ and energy $gE$ yields
eigenstates with quantum numbers $L+1,L_c+1,gE$. Similarly, the application of
$D_c$ to such eigenstates  yields eigenstates with quantum numbers 
$L-1,L_c-1,gE$. These
properties are well known for the case of a two-body contact interaction
\cite{Perelomov,Pitaevskii} and generalize to the Hamiltonian (\ref{ham}).
Note that the operators $A_0$ and $A_1$ are simply given by
\ba
\label{simple}
\hat{A}_0 &=& {1\over 2}N(N-1),\nonumber\\
\hat{A}_1 &=& N(\hat{L}-\hat{L}_c).
\ea

Applying the operators $\hat{A}_m$ onto the elementary symmetric 
polynomials yields
\be
\label{master1}
\hat{A}_m \,e_1 = 0 \qquad \mbox{for $m\ge 1$}
\ee
Thus, the one-dimensional space ${\cal V}_0={\rm span}\{e_1^L\}$ is an
invariant subspace of the Hamiltonian and therefore an eigenstate. Further,
\be
\label{master}
\hat{A}_m \,e_\lambda = 0 \qquad \mbox{for $m\ge 3$}.
\ee   
This immediately shows that the $(L-1)$-dimensional vector space
\be
\label{V1} 
{\cal V}_1={\rm span}\left\{e_{L-\lambda}\,e_1^\lambda: 
\quad\lambda=0,1,2,\ldots,L-2\right\}
\ee 
is annihilated by $A_m$ for $m\ge 3$. The space 
${\cal W}_1={\cal V}_0\cup {\cal V}_1$ is an 
invariant subspace of the Hamiltonian (\ref{ham}) since
\be
\label{A2}
\hat{A}_2 \,e_\lambda = 
2\lambda N \,e_\lambda -2(N-\lambda+1)\,e_1\,e_{\lambda-1},
\ee
and 
\ban
\hat{A}_1 \,e_\lambda = N\lambda\,e_\lambda - 
(N-\lambda+1)\,e_1\,e_{\lambda-1}.
\ean
Eigenfunctions of the Hamiltonian (\ref{ham}) in the subspace ${\cal W}_1$ are 
obtained by constructing
states with good $L_c$ quantum numbers. To this purpose we project $e_L$ onto
the space with $L_c=0$ and obtain
\ba
\label{ground}
e_L(z_1-z_c,\ldots,z_N-z_c) &=& \sum_{m=0}^L{1\over m!}\,(-z_c)^m\,D_c^m\, 
\, e_L(z_1,\ldots,z_N)\nonumber\\
&=& {(-1)^L\over L!}(\hat{L}_c-1)\,(\hat{L}_c-2)\ldots(\hat{L}_c-L)\, 
e_L(z_1,\ldots,z_N).
\ea
The r.h.s. of the first line is the Taylor expansion yielding a shift in the
single-particle coordinates by $-z_c$. The second line is obtained by combining
the operators $D_c$ and $z_c$ to the operator for the angular momentum of the 
center of mass (\ref{Lc}). This yields an operator that projects onto the 
subspace with $L_c=0$. The expansion \cite{PapBert}
\be
\label{expand}
e_L(z_1-z_c,\ldots,z_N-z_c) 
=\sum_{m=0}^L{N-m\choose L-m} (-z_c)^{L-m} e_m(z_1,\ldots,z_N).
\ee
shows explicitly that only states of ${\cal W}_1$ are involved in the 
construction of a state with $L_c=0$. The states 
\be
\label{special}
z_c^\lambda\,e_{L-\lambda}(z_1-z_c,\ldots,z_N-z_c),\qquad 
\lambda=0,1,2,\ldots,L-2,L
\ee
thus constitute an orthogonal basis of ${\cal W}_1$. Since they have 
different quantum
numbers $L_c=\lambda$ they are also eigenstates of the Hamiltonian
(\ref{ham}). Note that these wave functions do not depend on the coefficients
$c_m$ in the Hamiltonian. The corresponding energies are obtained by applying
the Hamiltonian (\ref{ham}) to the states (\ref{special}). Using
eqs. (\ref{simple}),(\ref{A2}) and (\ref{expand}) yields
\be
E_\lambda=
g\left[{1\over 2}N(N-1)c_0 + N(L-\lambda)(c_1+2c_2)\right],\qquad 
\lambda=0,1,2,\ldots,L-2,L. 
\ee
The states (\ref{special}) thus form a ladder. Let us consider the important
case of a two-body contact interaction. Analytical arguments \cite{Wilkin}
showed that the ground state is obtained for $\lambda=L$ in the case of
attractive interactions $g<0$. For repulsive interactions, a combination of
analytical \cite{Jackson2,Smith,PapBert} and numerical calculations
\cite{Bertsch} showed that the ground state is obtained for $\lambda=0$.
It is interesting to note that the corresponding fermionic problem, i.e. the
quantum Hall effect with zero range interaction potentials, also has an exact
ground state solution -- the Laughlin wave function\cite{Trugman}.

Note that the annihilation properties described in eq.~(\ref{master1}) and
eq.~(\ref{master}) may be generalized to further subspaces.
Let 
\be
\label{Vk}
{\cal V}_k={\rm span}
\left\{e_{\lambda_1}\,e_{\lambda_2}\ldots e_{\lambda_k}\,e_1^m: 
\quad m=0,1,\ldots,L-2k;\quad 
\sum_{j=1}^k \lambda_j =L-m;\quad\lambda_j>1\right\}
\ee
denote the space of wave functions that contain products of $k$ elementary 
polynomials different from $e_1$ and $e_0$. Using eq.~(\ref{master1}) and
eq.~(\ref{master}) one obtains
\be
\label{Agen}
\hat{A}_m\,{\cal V}_k = 0 \qquad \mbox{for $m\ge 2k+1$}.
\ee
However, these spaces are not invariant subspaces of the Hamiltonian, 
since e.g. $e_2^3(z_1,\ldots,z_6)\in \hat{V}e_3^2(z_1,\ldots,z_6)$.
Let us consider the space that is generated when acting with the
Hamiltonian (\ref{ham}) onto ${\cal V}_k$ for $k>1$. A direct calculation 
shows that
\ban
(\partial_i-\partial_j)^\mu\,(z_m-z_n)^\nu (\partial_m-\partial_n)^\nu 
\,{\cal V}_k =0 \qquad \mbox{for $\mu \ge \nu+2k+1$}.
\ean
Therefore, 
\ban
\label{annihil}
\hat{A}_m\,\hat{V} \,{\cal V}_k=0 \qquad \mbox{for $m \ge 4k+1$}.
\ean
Introducing the subspaces 
\ban
{\cal W}_k= \cup_{j=0}^k {\cal V}_j
\ean 
yields
\ban
\hat{V} \,{\cal W}_k \subset {\cal W}_{2k}\qquad\mbox{for $k>1$}.
\ean
The Hamiltonian matrix thus can be cast into a block-tridiagonal form.
The invariant subspaces ${\cal W}_0$, and ${\cal W}_1$ yield a block-diagonal
structure, and the remaining matrix is block-tridiagonal.

The knowledge of $L$ eigenstates (\ref{special}) permits us to determine
two additional eigenstates that are independent of the specific coefficients
in the Hamiltonian (\ref{ham}). 
For $L=4$ and $L=5$, there are two basis states with $L_c=0$.
One of them has already been determined above, i.e. 
$e_L(z_1-z_c,\ldots,z_N-z_c)$. The other two wave functions in question are
\ban
\phi_4&=& N\sum_{j=1}^N(z_j-z_c)^4
+3\left(\sum_{j=1}^N(z_j-z_c)^2\right)^2,\nonumber\\ 
\phi_5&=&
N\sum_{j=1}^N(z_j-z_c)^5 +2\sum_{j=1}^N(z_j-z_c)^2\,\sum_{i=1}^N(z_i-z_c)^3.
\ean
These wave functions are polynomials of degree four and five, respectively. 
Therefore, one finds $\phi_4,\phi_5\in {\cal W}_2$. 

The results derived so far were obtained for $0\le L\le N$. For $L>N$ the
Hilbert space depends on both $L$ and $N$. The definition of the spaces ${\cal
V}_k$ has to be modified correspondingly, e.g. ${\cal V}_1={\rm
span}\{e_{L-\lambda}\,e_1^\lambda:\,\,\lambda =\max{(0,L-N)},\ldots,L-2\}$ to
be compared with eq. (\ref{V1}).  However, the annihilation property described
in eq.~(\ref{Agen}) remains valid and the Hamiltonian matrix continues to be of
block-tridiagonal form. Note that the wave functions (\ref{special}) are
restricted to parameter values $\lambda=\max{(0,L-N)},\ldots,L-3,L-2,L$ for
$L>N$.

Quantum mechanical systems with invariant subspaces have received 
considerable attention in recent years, since
(quasi) exactly solvable systems fall into this class. For a general discussion
see, e.g.  refs.\cite{Turbiner,Shifman,Ushveridze,Calogero}. In difference to
the present work, the interest there is on {\it infinite}-dimensional Hilbert
spaces with a nested sequence of finite dimensional subspaces, and most
research has been limited to systems with a few degrees of freedom.  Note
finally that partially solvable many-body systems \cite{Calogero} or systems
with partial dynamical symmetry \cite{Leviatan,Ginocchio,Escher} also permit
the analytical computation of selected eigenstates.
 
In summary, we studied a large class of harmonically trapped, interacting boson
systems at angular momentum $L$. The Hamiltonian leaves a $L$-dimensional
subspace invariant, and the quantization of the angular momentum for the center
of mass yields several eigenstates that are independent of the details of the
two-body interaction. Important examples of suitable interactions are two-body
potentials of zero range and interactions that are analytically in the squared
inter-particle distance.

This research was supported by the U.S. Department of Energy under Grant
DE-FG03-00-ER41132.

\end{document}